\documentclass[a4paper,11pt]{article}
\usepackage{pos}

\title{Gradient tomography of jet quenching in high-energy heavy-ion collisions}
\author[a]{Yayun He}
\author[a]{Longgang Pang}
\author*[a,b]{Xin-Nian Wang}

\affiliation[a]{Key Laboratory of Quark \& Lepton Physics (MOE) and Institute of Particle Physics,\\
 Central China Normal University, Wuhan 430079, China}

\affiliation[b]{Nuclear Science Division, MS70R0319, Lawrence Berkeley National Laboratory, \\Berkeley, CA 94720, USA}

\emailAdd{xnwang@lbl.gov}
\abstract{
{\bf Abstract}: We illustrate with both a Boltzmann diffusion equation and full simulations of jet propagation  in heavy-ion collisions within the Linear Boltzmann Transport (LBT) model that the spatial gradient of the jet transport coefficient perpendicular to the propagation direction can lead to a drift and asymmetry in the transverse momentum distribution. Such an asymmetry depends on both the spatial position along the transverse gradience and  the propagating length. It can be used to localize the initial jet production positions for more detailed studies of jet quenching and properties of the quark-gluon plasma in heavy-ion collisions.

\section{Introduction} 
When a jet propagates through the medium and goes through multiple interaction under the eikonal approximation, we normally assume that it travels along a straight line and neglect the transverse diffusion. Under such an approximation, the scattering amplitude and the transverse momentum dependent quark distribution can be written as a Wilson line or gauge link along the trajectory between quark field operators \cite{Liang:2008vz},
\begin{equation}
f_A^q(x,\vec k_\perp)=\int(d y^-/4\pi) (d^2y_\perp/4\pi^2)
e^{ixp^+y^- -i \vec k_\perp\cdot \vec y_\perp} \langle A |\bar \psi(0) \gamma^+ {\cal L}(0,y)\psi(y) | A\rangle .
\end{equation}

Taking a Taylor expansion with respect to the transverse coordinate and assuming all higher-order gluon field correlators can be approximated as the product of the two-gluon field correlator, which is equivalent to dipole approximation at short distance, the transverse momentum distribution is found to satisfy a diffusion equation \cite{Liang:2008vz},
\begin{equation}
\partial_{\xi^-} f(\vec k_\perp)=(\hat q/4) \nabla_{k_\perp}^2 f(\vec k_\perp);
\;\; \hat q=(2\pi^2\alpha_{\rm s}/N_c) \rho xG(x)\mid_{x\approx 0}=\rho\hspace{-4 pt} \int \hspace{-3 pt} dk_T^2 k_\perp^2 (d\sigma/dk_T^2),
\end{equation}
with the diffusion constant given by the jet transport coefficients $\hat q$ which is the averaged transverse momentum broadening squared per unit length and proportional to the gluon density distribution inside the medium. In a static and uniform medium, the solution to the diffusion equation is a Gaussian with the width given by the total transverse momentum broadening squared $\hat q \xi^-$.
 }

\FullConference{%
  HardProbes2020, 1-6 June 2020, Austin, Texas}

\begin{document}
\maketitle

%\section{Introduction}

The jet transport efficient $\hat q$ also enters in both elastic and radiative energy loss for a propagating parton \cite{Baier:1996sk,Wang:2001ifa}.
Using the energy loss from elastic and radiative processes, one can calculate the energy loss distributions or the effective medium-induced splitting functions which one can use to calculate the final medium-modified jet fragmentation functions through the medium-modified DGLAP evolution equations \cite{Wang:2001ifa}. With these medium modified fragmentation functions one can then calculate the medium modified high $p_T$ hadron spectra and the suppression factors in heavy-ion collisions. One therefore can indirectly study medium properties through measurements of jet quenching or the suppression of leading hadrons.  The JET Collaboration recently has carried out a systematic analysis of the suppression of single inclusive hadron spectra in heavy-ion collisions at both RHIC and LHC energies and extracted the initial value of $\hat q$ at the center of the most central collisions \cite{Burke:2013yra}, assuming the space-time profile of the evolving medium is given by the relativistic hydrodynamics. The extracted values are two orders of magnitude larger than that in cold nuclei. 

\section{Gradience of the jet transport coefficient and transverse asymmetry}

If one neglects the inelastic scattering for a moment and considers the drag term from the elastic energy loss negligible, the Boltzmann transport equation can be cast into a drift-diffusion equation for the phase-space distribution of the propagating parton,
\begin{equation}
\partial_t f+(\vec k_\perp/E)\cdot \vec\nabla_{r_\perp} f=(\hat q/4) \nabla_{k_\perp}^2 f(\vec k_\perp,\vec r_\perp),
\end{equation}
where we assume the transverse momentum is much smaller than the longitudinal momentum $k_T/E\ll 1$.  Surprisingly we have found a simple but nontrivial  analytical solution to this drift-diffusion equation in a static and uniform medium \cite{He:2020iow},
\begin{equation}
f(\vec k_\perp,\vec r_\perp) =3(4E/\hat q t^2)^2 
e^{-(\vec{r}_\perp-\frac{\vec k_\perp}{2E}t)^2\frac{12E^2}{\hat q t^3}-\frac{k_\perp^2}{\hat q t}}.
\end{equation}
It clearly shows a drift in both space and transverse momentum, which is proportional to $\vec k_\perp$ and $\vec r_\perp$, respectively,  in addition to the diffusion in momentum and space.  
%The drift in momentum (space) is proportional to the spatial position (transverse momentum).
%while the diffusion in space and momentum are both determined by the jet transport coefficient.  
After integration over space (momentum) one can get the usual diffusion distribution in momentum (space).

 \begin{figure}
 \begin{center}
\begin{tabular}{l} 
\vspace{2.5 in}
\includegraphics[width=6.0cm]{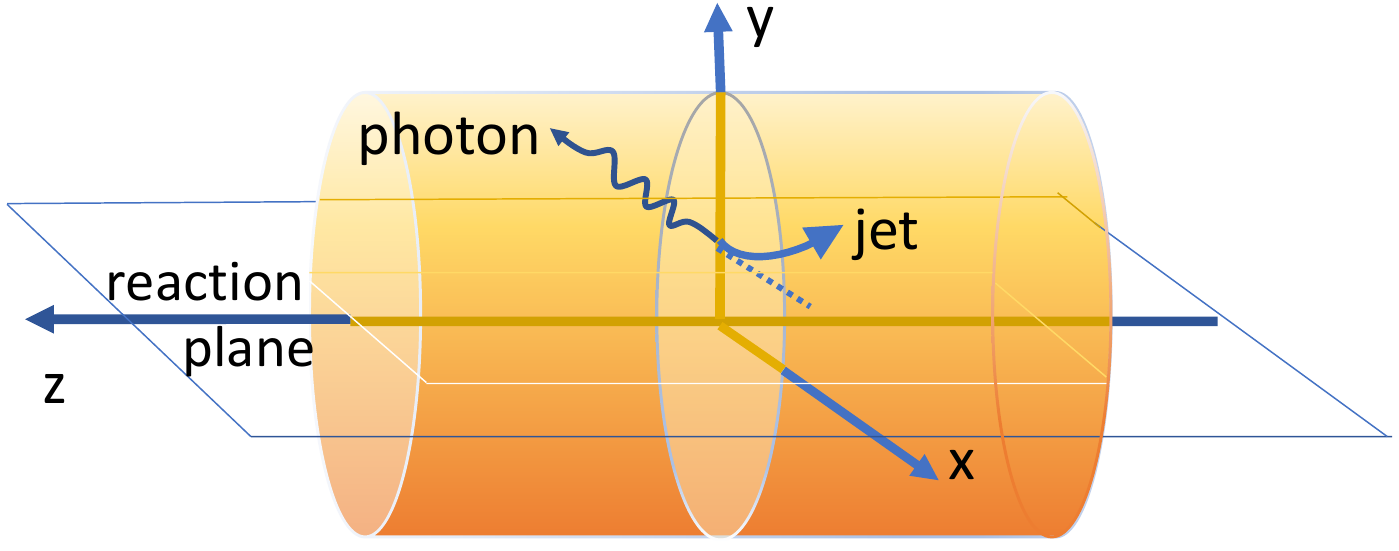}
\end{tabular}
\includegraphics[width=6.0 cm]{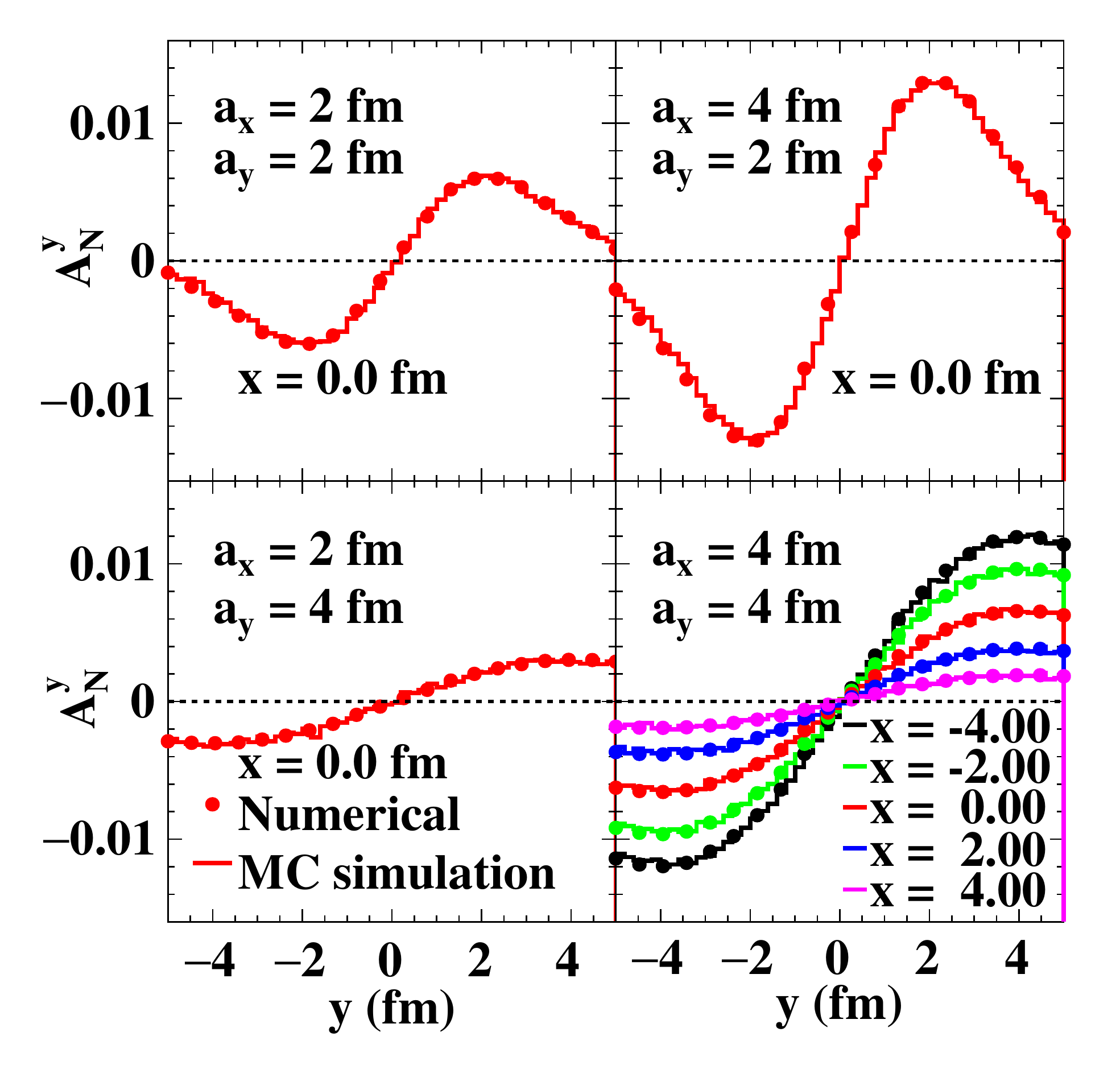}
\end{center}
\vspace{-2.0 in}
 \caption{(left) Illustration of a $\gamma$-jet and (right) transverse asymmetry as a function of the initial transverse position $y$ for different position $x$ along the path for $\hat q_0=5$ GeV$^2$/fm and $t_0=0.5$ fm/$c$ in the toy model.}
 \label{illustration}
\end{figure}

In a dynamic and nonuniform medium, one can solve the drift-diffusion equation either by numerical differentiation or Monte Carlo simulations. To investigate the momentum diffusion in the dynamic medium produced in heavy-ion collisions, we first set up a toy model for jet transport coefficient, $\hat q(\vec r_\perp,t)= \hat q_0 t_0e^{-x^2/a_x^2 - y^2/a_y^2}/(t_0+t)$, that mimics a longitudinally expanding medium with varying ellipticity and medium size through $a_x$ and $a_y$ in the transverse plane. Because of the transverse gradient of $\hat q$, one should expect a propagating parton to develop an asymmetry in its transverse momentum distribution along the direction of the gradience. For a parton, such as in the $\gamma$-jet event as illustrated in Fig.~1 (left), whose propagation direction relative to the event plane is known from the trigger, one can define the transverse asymmetry relative to the plane $\vec n=\hat {\vec p}_T \times \hat {\vec z}$ given by the beam $\hat {\vec z}$ and parton propagation direction $\hat {\vec p}_T$,
\begin{equation}
A_N^{\vec n}=\int d^3rd^3kf_a(\vec k,\vec r) {\rm Sign}(\vec k\cdot \vec n)\text{\LARGE $/$} \int d^3rd^3kf_a(\vec k,\vec x).
\end{equation}

Shown in Fig.~1 (right) are the transverse asymmetries as a function of the initial parton transverse position $y$ from the solution to the drift-diffusion equation via both numerical differentiation (close circles) and Monte Carlo simulations (histograms). We find indeed that the transverse asymmetry depends on the transverse position $y$ along the direction perpendicular to the plane $\vec n$. As the initial position gets away from the center ($y=0$), the gradient of $\hat q$ increases and so is the transverse asymmetry. As the initial position moves to the peripheral region of the dense matter at very large values of $y$, the gradience decreases again and so is the transverse asymmetry.  The asymmetry also increases with the propagation length (by varying the initial position $x$ along the propagation direction) for fixed initial transverse position $y$ (see the lower-right panel).  The asymmetry increases with the system size which increases the propagation length.   In noncentral collisions ($a_x\neq a_y$), if the parton propagates along the reaction plane ($a_x<a_y$) , the gradience and the asymmetry is smaller than when it propagates perpendicular to the reaction plane ($a_x>a_y$).  The asymmetry from the solution to the drift-diffusion equation for a toy model of a dynamic medium demonstrates well the principle of the gradient tomography.

\section{Gradient tomography of jet production}

\begin{figure}
\centerline{
\includegraphics[width=6.0cm]{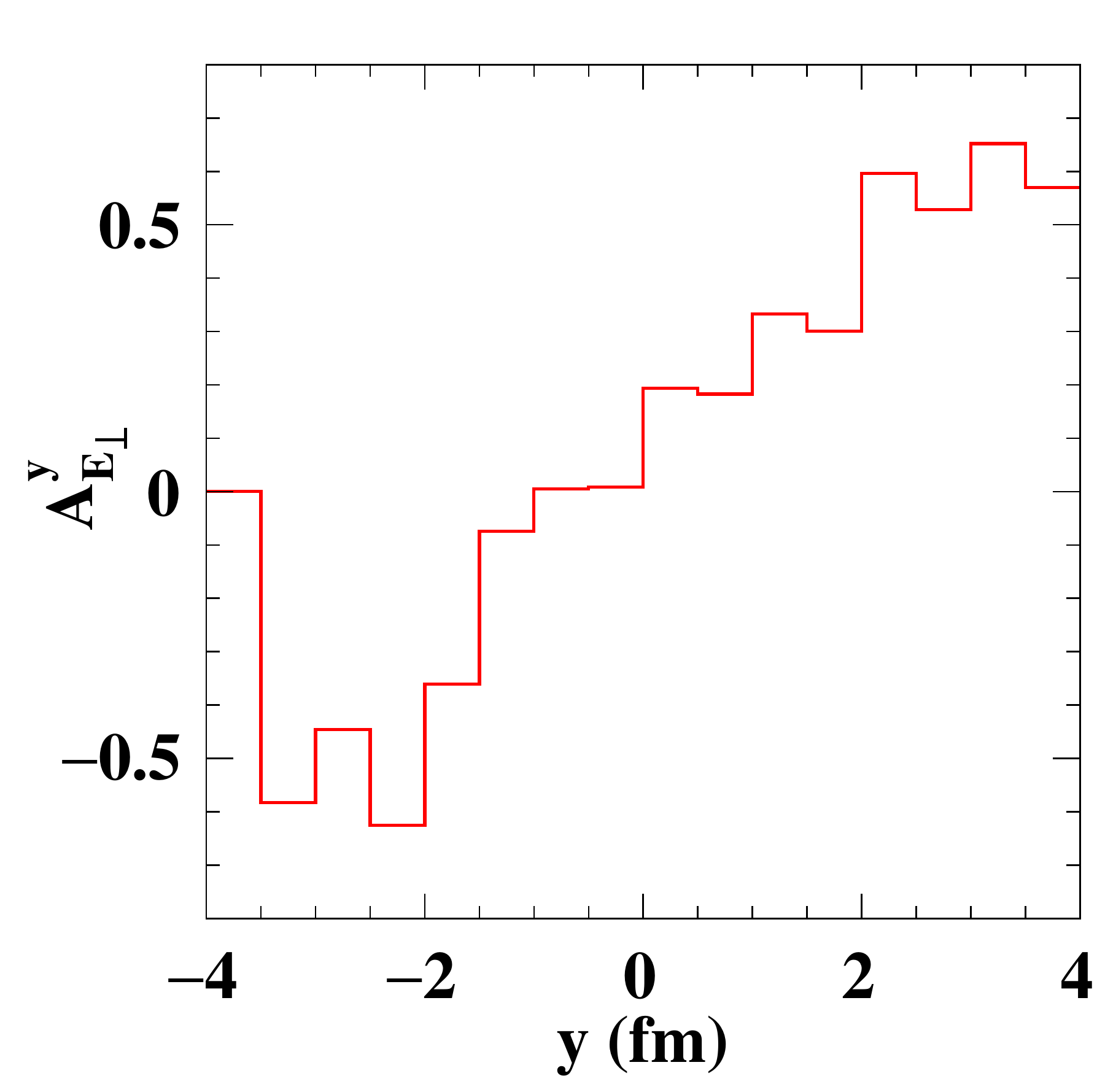}
\hspace{0.5in}
\includegraphics[width=6.5cm]{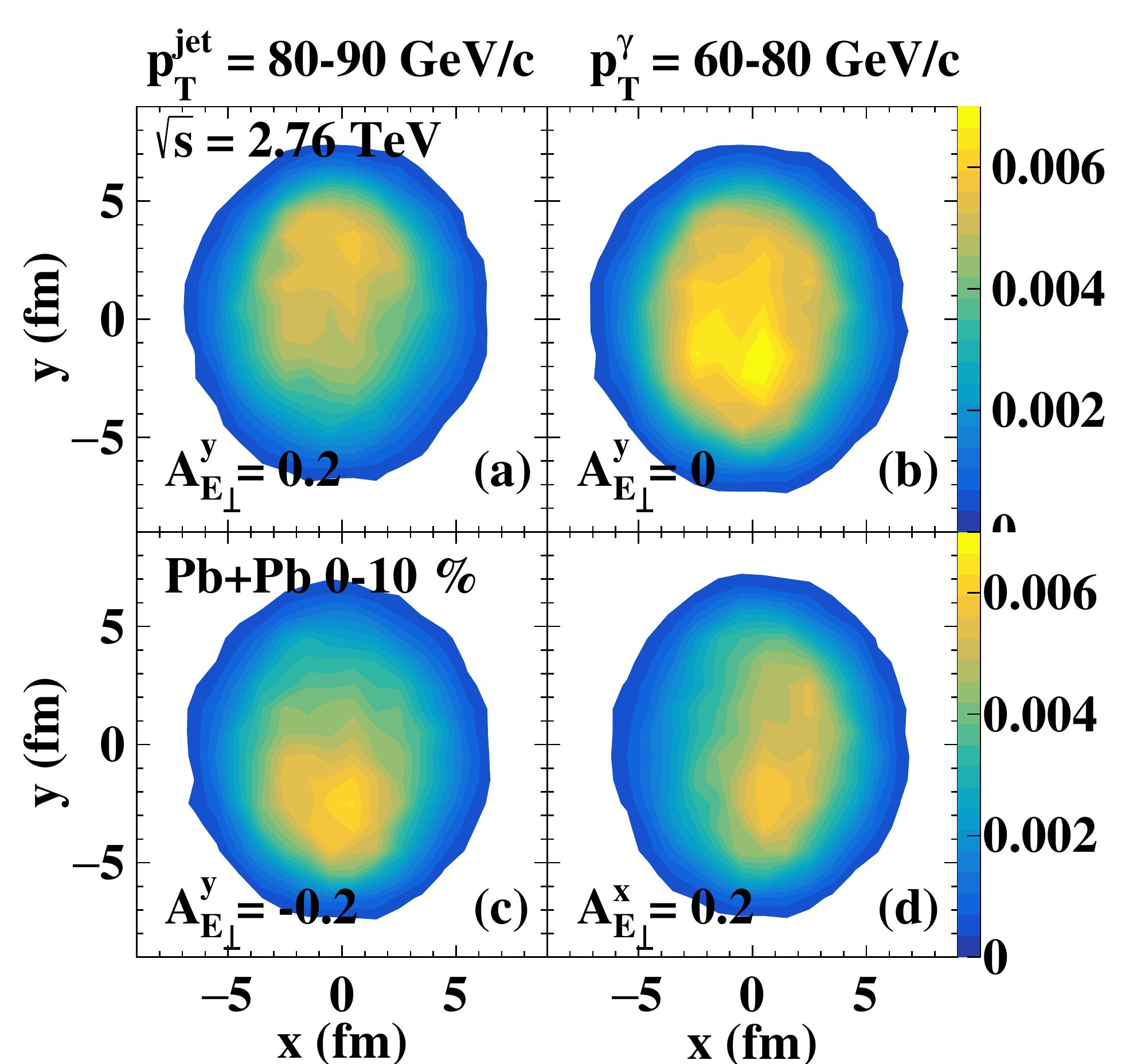}}
\vspace{-0.0 in}
 \caption{ (left) LBT results on the mean value of $A_{E_\perp}^y$ as a function of $y$ for $\gamma$-jets and (right) the transverse distribution of the initial $\gamma$-jet production position with a given value of transverse asymmetry  (a) $A_{E_\perp}^y=0.2$ (b) $A_{E_\perp}^y=0$ (c) $A_{E_\perp}^y=-0.2$, and (d) $A_{E_\perp}^x=0.2$ when the trigger photon is perpendicular to the event plane with $p_T^\gamma=60-80$ GeV/$c$ and $p_T^{\rm jet}=80-90$ GeV/$c$ in 0-10\% central Pb+Pb collisions at $\sqrt{s}=2.76$ TeV. }
 \label{x-y-asym}
\end{figure}

To illustrate the gradient tomography with the transverse asymmetry in a realistic medium in heavy-ion collisions, we use the linear Boltzmann transport (LBT) model \cite{He:2015pra,Luo:2018pto} for jet production and propagation, which includes both elastic and radiated processes for parton transport. The propagation and interaction of jet shower and recoil medium partons are simulated on an equal footing in a 3+1D  hydrodynamic medium of quark-gluon plasma (QGP).  LBT has been used successfully to describe the suppression of single inclusive hadron spectra, single inclusive jets and $\gamma$-jets \cite{Cao:2020wlm}.  We have confirmed the same picture as illustrated in Fig.~1 for the gradient tomography with the LBT simulations  \cite{He:2020iow}: a strong correlation between the transverse asymmetry and the transverse position of the initial jet production as shown in Fig.~2 (left).  By triggering on the value of the transverse asymmetry, one therefore should be able to localize the transverse position of the initial jet production. Shown in Fig.~2 (right), are the transverse distributions of the initial $\gamma$-jet production points with different values of the transverse asymmetry $A_{E_\perp}$ (weighted with the transverse momentum)  in 0-10\% central Pb+Pb collisions at $\sqrt{s}=2.76$ TeV.  One can see that for $A_{E_\perp}=0$ the distribution is concentrated at the center of the overlapped region of Pb+Pb collisions. When  $A_{E_\perp}$ is negative (positive) the distribution is shifted towards lower (upper) half of the plane.  If one changes the direction of the $\gamma$-jet from in-plane to out-plane (lower-right panel), one then biases the initial production points to a different region of the dense matter. Therefore, by triggering on the transverse asymmetry relative to the planed defined by the direction of the trigger in $\gamma$-jet, hadron-jet and dijet events, one can localize the initial production points of the jets. This will enable 
many more detailed studies of the properties of QGP in heavy-ion collisions, for example, the length dependence of jet energy loss or the spatial distribution of the jet transport coefficient. 

\vspace{8pt}

{\bf Acknowledgements}: This work is supported by DOE under Contract No. DE-AC02-05CH11231, NSF under Grant No. ACI-1550228 within the JETSCAPE Collaboration and NSFC under Grant Nos. 11935007, 11861131009 and 11890714. Computations are performed at DOE NERSC.

\end{document}